\documentclass{mn2e}

\usepackage{graphicx}
\newcommand{\citep}[1]{\cite{#1}}
\newcommand{\citet}[1]{\cite{#1}}

\newcommand{\bT}{\bmath{T}}

\newcommand{\bj}{\bmath{j}}
\newcommand{\bV}{\bmath{V}}
\newcommand{\bv}{\bmath{v}}
\newcommand{\tv}{\tilde{v}}

\newcommand{\bp}{\bmath{p}}
\newcommand{\br}{\bmath{r}}
\newcommand{\bg}{\bmath{g}}

\newcommand{\bB}{\bmath{B}}
\newcommand{\bb}{\bmath{b}}
\newcommand{\tb}{\tilde{b}}
\newcommand{\bE}{\bmath{E}}
\newcommand{\be}{\bmath{e}}
\newcommand{\cB}{{\cal B}}
\newcommand{\cE}{{\cal E}}

\newcommand{\bS}{\bmath{S}}

\newcommand{\dy}{\bar{y}}

\newcommand{\dv}{\bar{v}}
\newcommand{\db}{\bar{b}}
\newcommand{\dk}{\bar{k}}
\newcommand{\dw}{\bar{w}}

\newcommand{\text}[1]{\quad\mbox{#1}\quad}
\newcommand{\spr}[2]{\bmath{#1} \!\cdot\! \bmath{#2}}
\newcommand{\vpr}[2]{\bmath{#1} \!\times\! \bmath{#2}}
\newcommand{\vdiv}[1]{\spr{\nabla}{#1}}
\newcommand{\vcurl}[1]{\vpr{\nabla}{#1}}

\newcommand{\pder}[2]{\frac{\partial #1}{\partial #2}}

\newcommand{\ort}[1]{ \bmath{i}_{#1} }

\newcommand{\sech}{\mbox{sech}}

\newcommand{\beq}{\begin{equation}}
\newcommand{\ee}{\end{equation}}

\newcommand{\ba}{\begin{eqnarray}}
\newcommand{\ea}{\end{eqnarray}}
\newcommand{\om}{\omega}
\newcommand{\Alfven}{ Alfv\'{e}n }
\newcommand\etal{\textit{et al.\ }}
\newcommand\eg{\textit{e.g.\ }}

\newcommand{\Bf}{{magnetic field\,}}
\newcommand{\Bfs}{{magnetic fields\,}}
\newcommand{\NS}{{neutron star\,}}
\newcommand{\NSs}{{neutron stars\,}}

\begin{document}
\title{
Tearing instability in relativistic 
magnetically dominated plasmas}
\author[S.S.Komissarov]
{S.S.Komissarov$^1$, M.Barkov$^{1,2}$, M.Lyutikov$^{3}$ \\
$^1$ Department of Applied Mathematics, the University of Leeds,
Leeds, LS2 9GT, UK. e-mail: serguei@maths.leeds.ac.uk\\ 
$^2$ Space Research Institute, Moscow, Russia\\
$^3$ University of British Columbia, 6224 Agricultural Road,
Vancouver, BC, V6T 1Z1, Canada}
\date{Received/Accepted}
\maketitle
                                                                                                
\begin{abstract}
Many astrophysical sources of high energy emission, such as black hole
magnetospheres, superstrongly magnetized neutron stars (magnetars),
and probably relativistic jets in Active Galactic Nuclei and Gamma Ray
Bursts involve relativistically magnetically dominated plasma. In such
plasma the energy density of magnetic field greatly exceeds the
thermal and the rest mass energy density of particles.  Therefore the
magnetic field is the main reservoir of energy and its dissipation may
power the bursting emission from these sources, in close analogy to
Solar flares.  One of the principal dissipative instabilities that may
lead to release of magnetic energy is the tearing instability.  In
this paper we study, both analytically and numerically, the
development of tearing instability in relativistically
magnetically-dominated plasma using the framework of resistive
magnetodynamics. We confirm and elucidate the previously obtained
result on the growth rate of the tearing mode: the shortest growth
time is the same as in the case of classical non-relativistic MHD,
namely $\tau = \sqrt{\tau_a \tau_d}$ where $\tau_a$ is the
\Alfven crossing time and $\tau_d$ is the resistive time of a current
layer.
\end{abstract}

                                                                                                
\section{Introduction}
\label{intro}
Dissipation of magnetic energy may power high energy emission in a
variety of relativistic astrophysical phenomena, e.g. pulsar wind
nebulae \cite{Cor,Usov,LyK,KiS}, jets of active galactic nuclei
\cite{RoL,JLT}, gamma-ray bursts \cite{Dre,DrS}, and magnetars
\cite{ToD,Lyu03}.

Perhaps the most clear-cut case for magnetic dissipation is magnetars
-- \NSs with superstrong \Bfs, sometimes in
excess of quantum \Bf $B_Q=m_e^2 c^3/\hbar e = 4 \times 10^{13}$G.
(Observationally, magnetars are separated in two classes -- Anomalous
X-ray Pulsars (AXPs) and the Soft Gamma-ray Repeaters (SGRs) -- both
showing X-ray flares and quiescent X-ray emission.)  A number of
evidence suggest that processes that lead to the production of
magnetars X-ray flares (and possibly of the persistent emission) are
similar to those operating in the Solar corona. The bursting activity
of SGRs is strongly intermittent, showing a power law dependence of
the number of flares on their energy, $dN/dE \sim E^{\alpha}$, with
$\alpha=1.66$ and log-normal distribution of waiting times between the
flares \cite{gogus}, both being similar to Solar flares.  Giant flares
(GFs below), immense explosions releasing up to $10^{46}$ ergs in a
fraction of a second, provide a number of independent evidence in
favour of magnetic dissipation.  Two recently observed explosions,
from SGR 1900+14 and SGR 1806-20, showed similar behavior leading to
and following the GFs: months before the GF X-ray activity increased,
spectrum hardened and spindown increased \citep{turolla}. In the
post-flare period, pulsed fraction and the spin-down rate have
significantly decreased and the spectrum softened
\citep{woods01,rea05}.  All these effects are in agreement with the
prediction of twisted magnetosphere model \citep{tlk02}, with the
twist increasing before the GF and decreasing during the GF brought
about by reconnection \cite{Lyu06}. Most importantly, observations of
the December 27 GF from SGR 1806-20 show very short rise time $\sim
0.25 $ msec \citep{palmer05}. Such short rise times are inconsistent
with crustal deformations and points to magnetospheric origin of GF
\cite{Lyu06}.

In analogy with Solar flares magnetic energy to be dissipated may
 build-up on long time scales and then be dissipated on very short
 times scales, possibly as short as \Alfven crossing time.  For
 example, a slow plastic motion of the crust implants a twist
 (current) in the magnetosphere on a long time scale. At some point a
 global system of magnetospheric currents and sheared magnetic fields
 loses equilibrium and produces a flare. This instability may be
 dynamical (\eg loss of magnetic equilibrium) and/or resistive (this
 work).

Though similar in underlying physical process, magnetar and Solar
plasmas may differ considerably, since properties of plasma expected
in magnetar magnetospheres are very different from the conventional
Solar and laboratory plasmas.  The principal difference is that
magnetar plasma is strongly (relativistically strongly) magnetized.
It is convenient to introduce a parameter $\sigma_m$ \citep{kc84} as
the ratio of the magnetic energy density $u_B=B^2/8\pi$ to the total
plasma energy density (including rest mass!): \beq
\label{sigmadef}
\sigma_m={u_B\over u_{p}}
\label{sig}
\ee Conventionally, in a non-relativistic Solar and laboratory
plasmas, parameter $\sigma_m$ is small, $\sigma_m \ll 1$. On the other
hand, in magnetar magnetospheres it is expected that $\sigma_m$ is
very large, \beq {\om_B R_{NS} \over c} \left({m_e \over m_p} \right)
\sim 10^{13} \leq \sigma_m \leq {\om_B \over \Omega} \left({m_e \over
m_p} \right) \sim 10^{16} \ee (the upper limit here comes from
electron-ion plasma density equal to the Goldreich-Julian density, the
lower limit corresponds to plasma density $\sim B/ (R_{NS} e)$ at
which point currents drifting with near the speed of light create
toroidal \Bfs of the order of the poloidal).  Here $\om_B = e B/m_e c$
is cyclotron frequency, $B$ is \Bf at the \NS surface, $R_{NS} $ is
radius of \NS and $\Omega$ is rotational frequency.

Large expected value of $\sigma_m$ (or small $1/\sigma_m$) may be used
as an expansion parameter in equations of relativistic
magnetohydrodynamics (MHD). The zero order equations, the equations of
{\it magnetodynamics} (MD), describe the dynamics of magnetic field
under the action of magnetic pressure and tension \cite{Kom02}. Since
to this order the rest mass density and thermodynamic pressure of
plasma completely vanish, the only effect plasma has on this dynamics
is due to its high conductivity which requires ${\bf E} \cdot {\bf
B}=0$ and $E^2<B^2$.  Alternatively, one can write the equations of
magnetodynamics as Maxwell's equations closed with a suitable Ohm's
law that gives the electric current as a function of the
electromagnetic field only \cite{Gru99}. This Ohm's law is derived
from the condition of vanishing Lorentz force that explains the
alternative name for this system, {\it force-free degenerate
electrodynamics} or simply {\it force-free electrodynamics}.

The limit $\sigma_m \rightarrow \infty$ is somewhat reminiscent of the
subsonic incompressible hydrodynamic. In this limit the \Alfven
four-velocity is $u_A \sim \sqrt{2 \sigma_m} \rightarrow \infty$, so
that both cases are applicable when (four)-velocities are much smaller
than the velocity of propagation of disturbances in a medium (\Alfven
and sound waves correspondingly).  In fact, this analogy turns out to
be much deeper: as we discuss in this paper there are deep
similarities in governing equations between magnetodynamics a small
velocity limit of the conventional magnetohydrodynamics.

In order to address energy dissipation and eventually particle
acceleration, one needs to go beyond the ideal approximation.  One
possible approach is to take into account the dissipation of mostly
force-free currents due to plasma resistivity.  Resistivity will
result in the decay of currents supporting the magnetic field, which
would influence the plasma dynamics.  Though plasma resistivity will
be of the anomalous type, mediated by plasma turbulence and not by
particle-particle collisions, as a first step we assume that plasma
resistivity can be represented by macroscopic resistivity parameter
$\eta$.  As another simplification we neglect a possible back reaction
of the heated plasma on the global dynamics. This may be justified if
the dissipated energy is quickly radiated away.

The principal resistive instability in a conventional,
non-relativistic plasma is a so-called tearing mode. It is one of the
principle unstable resistive modes, which plays the main role in
various TOKAMAK discharges like sawtooth oscillations and major
disruptions \citep{Kadomtsev75}, unsteady reconnection in Solar flares
\citep{Shivamoggi85,Aschwanden02} and Earth magnetotail \citep{gca78}.
Qualitatively, the most important property of the tearing mode is that
a current layer of thickness $l$ may dissipate on time scales much
shorter than it the resistive time scale $\tau_d \sim l^2/\eta$. In
addition, tearing mode may be an initial stage of the development of
the (steady-state) reconnection layers.

Development of the tearing mode in a relativistic, strongly magnetized
plasma ($\sigma_m \gg 1$) is the principal topic of this work.
Initially, this problem has been considered by Lyutikov
\shortcite{Lyu03}, who found that in resistive magnetodynamics the
current layers are unstable towards formation of resistive small-scale
current sheets, that's to development of tearing mode.  He also found
that the growth rate of tearing mode is intermediate between the short
\Alfven time scale $ \tau_a \sim l/v_a$ (which in a $\sigma_m \gg 1$
plasma is the light crossing time scale $\tau_c \sim l/c$) and a long
resistive time scale $\tau_d$: $\tau \sim (\tau_d
\tau_a)^{1/2}$. Surprisingly, this is exactly the same expression as
the one found in the non-relativistic framework of incompressible
magnetohydrodynamics.  In this paper we uncover the deep underlying
reasons for this coincidence. It turns out that slow,
resistively-driven evolution of strongly magnetized plasma is
described by a system of equations that is very similar to
nonrelativistic MHD.  In addition to the analytical study we test the
theory by means of numerical simulations.

In Section \ref{basics} we describe the basic equations of resistive
magnetodynamics.  In Section \ref{equi-approximation} we consider the
case of slow evolution near equilibrium and show that in this regime
the MD equations can be reduced to the system that is very similar to
nonrelativistic MHD.  Further reduction is possible when the
equilibrium is characterized by zero magnetic tension. This is shown
in Section \ref{incomp-limit}. The analytic theory of the tearing
instability is reviewed in Section \ref{tearing}. Section \ref{method}
describes our numerical method and the results of numerical
simulations are presented in Section \ref{simulations}.

\section{Basic equations}
\label{basics}

In any inertial frame of special relativity the dynamics of
electromagnetic field is described by Maxwell's equations.  They are
the Faraday law
\begin{equation}
\frac{1}{c}\pder{\bB}{t}+\vcurl{E} = 0,
\label{Max1}
\end{equation}

\noindent
the magnetic Gauss law
\begin{equation}
\vdiv{B}=0,
\label{Max2}
\end{equation}

\noindent
the Ampere law
\begin{equation}
-\frac{1}{c}\pder{\bE}{t}+\vcurl{B} = \frac{4\pi}{c} \bj,
\label{Max3}
\end{equation}

\noindent
and the electric Gauss law
\begin{equation}
\vdiv{E} = 4\pi\varrho .
\label{Max4}
\end{equation}
When these equations are supplemented with the Ohm law, that relates
the electric current with the electric field, the system become
closed. However, one can also write down additional equations that
describe the evolution of the energy and momentum of the
electromagnetic field. These additional equations are very useful for
physical interpretation of electromagnetic phenomena.

\noindent
Energy conservation law:

\begin{equation}
\pder{e}{t}+\vdiv{S}=-\spr{E}{j},
\label{E-cons}
\end{equation}
where
\begin{equation}
  e = \frac{E^2+B^2}{8\pi},
\label{E-dens}
\end{equation} 
is the energy density and
\begin{equation}
  \bS = \frac{c}{4\pi}\vpr{E}{B}
\label{E-flux}
\end{equation}
is the energy flux density (the Poynting flux).

\noindent
Momentum conservation law:
\begin{equation}
\pder{\bp}{t}+\vdiv{T}=-\varrho\bE -\frac{1}{c}\vpr{j}{B},
\label{p-cons}
\end{equation}
where
\begin{equation}
  \bp = \frac{1}{4\pi c} \vpr{E}{B},
\label{p-dens}
\end{equation}
is the momentum density and
\begin{equation}
  \bT = -\frac{1}{4\pi} (\bE\otimes\bE + \bB\otimes\bB)
        +\frac{1}{8\pi}(E^2+B^2)\bg
\label{T}
\end{equation}
is the Maxwell stress tensor. In the last equation $\bg$ is the metric
tensor of Euclidean space,

In strong magnetic field the conductivity across magnetic field can
become highly suppressed (when the Larmor frequency becomes much
higher than the frequency of particle collisions).  Moreover, if the
magnetic field is so strong that one can ignore the inertia of plasma
particles then the Ohm's law can be written as

\begin{equation}
\bj = \varrho \frac{\vpr{E}{B}}{B^2}c + \sigma_\parallel
\bE_\parallel,
\label{Ohm}
\end{equation}
where $E_\parallel$ is the component of electric field that is
parallel to the magnetic field.  Another condition for eq.(\ref{Ohm})
to hold is $E<B$. If this condition is not satisfied the cross-field
conductivity has also to be taken into account, e.g. via adding the
$\sigma_\perp \bE_\perp$ term to the right hand side of eq.\ref{Ohm}.
In general, the conductivity $\sigma_\parallel$, mediated by
particle-wave-particle interaction, should depend on the Lorentz
factor of plasma through Lorentz transformation of the effective
collision rates \cite{Lyu03}.  As a simplification, applicable in
cases when no strongly relativistic motion {\it along} the \Bf lines
are expected, here we will assume that $\sigma_\parallel$ is a
constant macroscopic parameter.

 A characteristic speeds of magnetized plasma is the drift speed of
charged particles across \Bf

\begin{equation}
   \bV = c\frac{\vpr{E}{B}}{B^2}.
\label{V-drift}
\end{equation}
Note that it appears in the Ohm's law (\ref{Ohm}) where it describes
the non-conductive contribution to the current density.

Given the Ohm law (\ref{Ohm}) one can write the energy and momentum
conservation laws as

\begin{equation}
\pder{e}{t}+\vdiv{S}=-\sigma_\parallel E_\parallel^2,
\label{E-cons-1}
\end{equation}

\begin{equation}
\pder{\bp}{t}+\vdiv{T}=-\varrho\bE_\parallel,
\label{p-cons-1}
\end{equation}
In the limit of infinite conductivity, $\sigma_\parallel \to \infty$,
the electric current $j_\parallel=\sigma_\parallel E_\parallel$ must
remain finite and this ensures $\bE_\parallel \propto
\sigma_\parallel^{-1} \to 0$.  In this limit the dynamics of the
electromagnetic field can be described by the following closed set of
equations \cite{Kom02}

\begin{equation}
\frac{1}{c}\pder{\bB}{t}+\vcurl{E} = 0,
\label{MD1}
\end{equation}
                                                                                                                  
\begin{equation}
\vdiv{B}=0,
\label{MD2}
\end{equation}

\begin{equation}
\pder{e}{t}+\vdiv{S}=0,
\label{MD3}
\end{equation}
                                                                                                                  
\begin{equation}
\pder{\bp}{t}+\vdiv{T}=0.
\label{MD4}
\end{equation}
In \cite{Kom02} this system of equations was derived from the system
of ideal relativistic MHD in the limit of vanishing rest mass density
and pressure of matter. This way of derivation suggests to call this
system {\it magnetodynamics} (MD), the name that is obtained from {\it
magnetohydrodynamics} (MHD) via excluding its {\it hydro} part.
Alternatively, it could be called {\it force-free degenerate
electrodynamics}. This name originates from the early attempts to
construct steady-state models of magnetospheres satisfying the
degeneracy condition, $$\spr{E}{B}=0,$$ and the force-free condition,
$$\varrho\bE +\frac{1}{c}\vpr{j}{B}=0, $$ e.g. \cite{TM}.  However, it
is longer and makes an unwarranted emphasis on the electric component
of the electromagnetic field.  Indeed, similarly to ideal MHD where
the electric field vanishes in the fluid frame, in ideal MD the
electric field vanishes in the frame moving with the drift velocity.
Thus, the electric component of the field can still be considered as a
secondary one that is induced by the motion of magnetized plasma.  In
fact the electric field vector can be eliminated from the set of
dependent variables of equations (\ref{MD1}-\ref{MD4}) via

\begin{equation}
\bE = -\frac{1}{c} \vpr{V}{B},
\label{V1}
\end{equation}

\begin{equation}
  e = \frac{B^2}{8\pi} \left(1+\frac{V^2}{c^2} \right),
\label{E-dens-a}
\end{equation}

\begin{equation}
  \bS = \frac{B^2}{4\pi} \bV,
\label{Poynting-a}
\end{equation}

\begin{equation}
  \bp = \frac{B^2}{4\pi c^2} \bV,
\label{p-dens-a}
\end{equation}

\begin{equation}
  \bT = -\frac{1}{4\pi} \left[ \frac{1}{c^2}
       (\vpr{V}{B})\otimes(\vpr{V}{B}) + \bB\otimes\bB\right] +
\label{T-a}
\end{equation}
$$ +\frac{1}{8\pi}B^2 \left(1+\frac{V^2}{c^2}\right) \bg.
$$ Moreover, in the dissipative regime the electromagnetic field can
no longer be described as force-free either (see eq.\ref{p-cons-1} ).

\section{The quasi-equilibrium approximation.}
\label{equi-approximation}

Here we will derive equations that describe slow evolution of strong
electromagnetic field near the state of force-free equilibrium.  There
are two basic characteristic time scales in such problems, the light
crossing time scale
\begin{equation}
   \tau_c = l/c
\end{equation}
and the diffusion time scale
\begin{equation}
    \tau_d = l^2/\eta,
\end{equation}
where $\eta =c^2/4\pi\sigma_\parallel$ is the resistivity and $l$ is
the characteristic length scale of the problem.  Since in the limit of
magnetodynamics the hyperbolic waves (fast and Alfv\'en waves)
propagate with the speed of light then $\tau_c$ is the typical time of
establishing dynamical equilibrium. $\tau_d$ gives the time scale for
the diffusion of magnetic field due to finite resistivity.  For
systems close to a static equilibrium the characteristic time scale of
their global evolution, which we will denote as $\tau$, is much larger
than $\tau_c$. This gives us the small parameter
\begin{equation}
   \mu_c = \tau_c/\tau \ll 1 .
\end{equation}
Diffusion is usually only important on small scales where there may
exist large gradients of physical quantities. Therefore we will also
assume that $\tau$ is much smaller than $\tau_d$ and this gives us
another small parameter
\begin{equation}
   \mu_d = \tau/\tau_d \ll 1.
\end{equation}
This automatically ensures that the relativistic Lundquist number
\begin{equation}
    L_u = \sqrt{2} \tau_d / \tau_c = \sqrt{2}lc/\eta \gg 1
\end{equation}
($\sqrt{2}$ is introduced in order to simplify the expressions of 
Sec.\ref{tearing}.)
Moreover, from the Faraday equation one immediately obtains
\begin{equation}
    \cE/\cB = \mu_c \ll 1
\label{EllB}
\end{equation}
where $\cE$ and $\cB$ are the characteristic scales of the electric
and the magnetic fields respectively. Thus, the electric field is much
weaker than the magnetic field and this results in the drift speed
much smaller than the speed of light (see eq.\ref{V-drift})

The Ohm law (\ref{Ohm}) leads to the following dimensionless form of
the Ampere equation

\begin{equation}
-\mu_c^2\pder{\bE}{t}+\vcurl{B} = \mu_c^2 \varrho
    \frac{\vpr{E}{B}}{B^2} + \delta_1 \bE_\parallel,
\label{Amp}
\end{equation}
where
\begin{equation}
 \delta_1=(4\pi)^2 L_u \frac{\cE_\parallel}{\cB}.
\end{equation}
Thus, the Ampere law may be reduced to
\begin{equation}
\vcurl{B} = \frac{4\pi}{c} \sigma_\parallel \bE_\parallel
\label{Amp-1}
\end{equation}
and the characteristic scale for the parallel component of the
electric field is
\begin{equation}
\cE_\parallel = (4\pi)^{-2}L_u^{-1} \cB \ll \cB.
\label{E-para}
\end{equation}
Using this equation and the expression (\ref{V-drift}) for the drift
velocity we can write
\begin{equation}
  \bE = -\frac{1}{c}\vpr{V}{B} + \frac{c}{4\pi \sigma_\parallel}
  \vcurl{B}.
\label{E-tot}
\end{equation}
Substitution of this result into the Faraday equation leads to the
exactly the same advection-diffusion equation for magnetic field as in
the nonrelativistic MHD

\begin{equation}
  \pder{\bB}{t} -\nabla\times(\vpr{V}{B}) - \eta \nabla^2{\bB}=0.
\label{Adv-Diff}
\end{equation}
  
The dimensionless form of the energy equation (\ref{E-cons-1}) reads
\begin{equation}
\pder{e}{t}+\vdiv{S}=-\delta_2 E_\parallel^2,
\end{equation}
where
\begin{equation}
     \delta_2 = \frac{\mu_d}{(4\pi)^2} \ll 1 .
\end{equation}
Thus, the dissipative term may be omitted and the energy equation
reduces to its ideal form
                                                                                          
\begin{equation}
\pder{e}{t}+\vdiv{S}=0 .
\label{E-cons-2}
\end{equation}
The dimensionless form of the momentum equation (\ref{p-cons-1}) is

\begin{equation}
\mu_c^2 \pder{\bp}{t}+\vdiv{T}=-\delta_3 \rho E_\parallel,
\end{equation}
where
\begin{equation}
  \frac{\delta_3}{\mu_c^2}=\frac{\mu_d}{4\pi} \ll 1.
\end{equation}
Thus, the resistive term in the momentum equation can also be omitted
and we write

\begin{equation}
\pder{\bp}{t}+\vdiv{T}=0.
\label{p-cons-2}
\end{equation}

\noindent
Given condition (\ref{EllB}) we may write
\begin{equation}
   e=\frac{B^2}{8\pi}
\end{equation}
and
\begin{equation}
  \bT = -\frac{1}{4\pi}\bB\otimes\bB +\bg \frac{B^2}{8\pi}.
\end{equation}
Then we combine eq.(\ref{Max2}), eq.(\ref{Adv-Diff}),
(\ref{E-cons-2}), and (\ref{p-cons-2}) into a closed system of
equations very similar to nonrelativistic resistive MHD

\begin{equation}
\vdiv{B}=0,
\end{equation}

\begin{equation}
  \pder{\bB}{t} -\nabla\times(\vpr{V}{B}) - \eta \nabla^2{\bB}=0.
\end{equation}

\begin{equation}
\pder{\rho}{t}+\nabla(2\rho\bV)=0,
\label{cont}
\end{equation}

\begin{equation}
\pder{\rho\bV}{t} + \nabla\left( -\frac{\bB\otimes\bB}{4\pi} +\bg
\frac{B^2}{8\pi}\right) =0.
\label{mom}
\end{equation}
In these equations
\begin{equation}
\rho=e/c^2 = \frac{B^2}{8\pi c^2}
\label{rho}
\end{equation}
is the mass density of the electromagnetic field. There are however
several differences. First, there is the factor of $2$ in the second
term of the ``continuity equation'' (\ref{cont}). Second, there is no
$\nabla(\rho\bV\otimes\bV)$ term in the momentum equation. Finally,
there is no ``energy equation''. To some extent this similarity has
been noticed in \cite{Gru99}.

\section{The incompressible limit.}
\label{incomp-limit}

This system can be reduced even further if the equilibrium is
supported predominantly by the magnetic pressure.  Indeed, the
dimensionless equations of momentum and continuity include small
parameter $\epsilon=\mu_c^2$

\begin{equation}
  \epsilon \pder{\rho\bV}{t} - \frac{1}{4\pi} (\spr{B}{\nabla}) \bB
  +\nabla \frac{B^2}{8\pi} =0.
\label{mom1}
\end{equation}
                                                                                
\begin{equation}
\pder{\rho}{t}+2\nabla(\rho\bV)=0,
\label{cont1}
\end{equation}
                                                                                
\noindent
Using the method of perturbation we expand the unknowns in powers of
$\epsilon$
$$ \bB=\bB_0+\epsilon\bB_1, \quad \bV=\bV_0+\epsilon\bV_1, \quad
\rho=\rho_0 +\epsilon\rho_1,
$$ where
\begin{equation}
 (\spr{B_0}{\nabla}) \bB_0=0.
\label{a0}
\end{equation}
                                                                                
\noindent
To the zero order we have
\noindent
\begin{equation}
    \rho_o=\frac{B_0^2}{8\pi c^2} = \mbox{const}
\label{a1}
\end{equation}
and
\begin{equation}
    \spr{\nabla}{\bV_0}=0,
\label{a2}
\end{equation}
\noindent
whereas the first order gives us
\begin{equation}
   \rho_0\pder{\bV_0}{t} - \frac{1}{4\pi} (\spr{B_1}{\nabla} \bB_0 +
  \spr{B_0}{\nabla} \bB_1) +\nabla \frac{\spr{B_1}{B_0}}{4\pi} =0.
\label{a3}
\end{equation}
This result shows us that eq.(\ref{mom1}) differs from
                                                                                
\begin{equation}
  \epsilon \rho \pder{\bV}{t} - \frac{1}{4\pi} (\spr{B}{\nabla}) \bB
  +\nabla \frac{B^2}{8\pi} =0
\label{mom2}
\end{equation}
only by terms of order $\epsilon^2$ and we may approximate equation
(\ref{mom}) by
\begin{equation}
   \rho \pder{\bV}{t} -\frac{1}{4\pi} (\spr{B}{\nabla}) \bB +\nabla
  \frac{B^2}{8\pi} =0.
\label{mom3}
\end{equation}
that implies the vorticity equation
\begin{equation}
   \rho \pder{\vcurl{V}}{t} -\frac{1}{4\pi} \nabla\times
  (\spr{B}{\nabla})\bB =0.
\label{mom4}
\end{equation}
 
\noindent
Thus, we have arrived to the following system of equations
                                                                                
\begin{equation}
\vdiv{B}=0,
\label{divb}
\end{equation}
                                                                                
\begin{equation}
    \spr{\nabla}{\bV}=0,
\label{divv}
\end{equation}
                                                                                
\begin{equation}
  \pder{\bB}{t} -\nabla\times(\vpr{V}{B}) - \eta \nabla^2{\bB}=0,
\label{induction}
\end{equation}
                                                                                
\begin{equation}
   \rho \pder{\vcurl{V}}{t} -\frac{1}{4\pi}
   \nabla\times(\spr{B}{\nabla})\bB =0.
\label{vorticity}
\end{equation}
 
\noindent
This system of equations is very similar to the one of incompressible
MHD. The only difference is that equation (\ref{mom3}) involves the
Eulerian time derivative whereas the vorticity equation of
incompressible MHD involves the Lagrangian time derivative (this,
however, has no effect on the perturbation equations of the linear
stability theory).  One may argue that the limit of incompressible
magnetodynamics is applicable only to a very limited range of problems
as in general case the magnetic tension cannot be ignored and as the
result one cannot assume that $\rho$ is constant.  We only point out
that exactly the same argument applies to incompressible MHD unless
the plasma magnetization is very low \cite{Biskamp}.

Both in the nonrelativistic case \cite{Priest} and in the case of
magnetically dominated relativistic plasma \cite{Lyu03} the most basic
equilibrium current sheet configuration involves 1-dimensional current
sheet with slab symmetry. Within this current sheet the magnetic field
gradually rotates so that its pressure remains constant and its
tension vanishes. Studying the stability of the relativistic current
sheet Lyutikov\shortcite{Lyu03} derived the same growth rates for the
tearing mode as in the nonrelativistic theory. Our analysis of the
equations of resistive magnetodynamics explains this result as the
direct consequence of the similarity between magnetodynamics and
non-relativistic MHD. In the following section we study the tearing
mode in some details.

\section{Tearing mode instability}
\label{tearing}

Consider the stationary versions of
eqs.(\ref{induction},\ref{vorticity})

\begin{equation}
  \nabla\times(\vpr{V}{B}) - \eta \nabla^2{\bB}=0,
\label{ind1}
\end{equation}
                                                                                
\begin{equation}
  \frac{1}{4\pi} \nabla\times(\spr{B}{\nabla})\bB =0.
\label{vort1}
\end{equation}
Low \shortcite{Low} found the following one-dimensional solution for
these equations

\begin{eqnarray}
\bB &=& B_0 \tanh(y/l)\ort{x} \pm B_0\sech(y/l)\ort{z} ,\\
\label{b-steady}
\bV &=& -(\eta /l)\tanh(y/l)\ort{y}.
\label{v-steady}
\end{eqnarray}
It describes a current sheet in which the magnetic field rotates {\it
exactly} by $\pi$ without changing its magnitude. The rotation
continues from $y=-\infty$ to $y=+\infty$ but most of it occurs within
$y\in[-l,l]$. This allows us to describe $l$ as the thickness of
current sheet. The total electric field ,
$\bE=-\vpr{V}{B}+\eta\vcurl{B}$, corresponding to this solution is

\begin{eqnarray}
  \bE = -\frac{B_0\eta}{l}\ort{z}.
\label{e-steady}
\end{eqnarray}
It turns out that the electric and magnetic fields given by equations
(\ref{b-steady},\ref{e-steady}) also satisfy the full set of
stationary Maxwell's equations supplemented with Ohm's law
(\ref{Ohm}). (As a matter of fact this provides us with a very good
test problem for our numerical code.)

Consider a perturbation of this equilibrium state of the form
$$ \bb(x,y,t) = \tilde{\bb}(y) \exp(i(k_xx+k_yy)+wt),
$$
$$ \bv(x,y,t) = \tilde{\bv}(y) \exp(i(k_xx+k_yy)+wt).
$$ Equation (\ref{induction}) and curl of equation (\ref{vorticity})
give us the following linearized equations for the the y-components of
the perturbation
                                                                                                
\begin{equation}
  w \tb_y = i \tv_y (\spr{k}{B}) + \eta (\tb_y^{''}-k^2\tb_y ),
\label{ind-lin}
\end{equation}
                                                                                                
\begin{equation}
  w(\tv_y^{''}-k^2\tv_y)= \frac{i(\spr{k}{B})}{4\pi\rho} \left[-\tb_y
  \frac{\spr{k}{B^{''}}}{\spr{k}{B}} + (\tb_y^{''}-k^2\tb_y)\right].
\label{vort-lin}
\end{equation}
These equations are the same as equations 6.6 and 6.7 in Priest \&
Forbes \shortcite{Priest} where the tearing mode instability is
studied in the non-relativistic framework of incompressible MHD
(Priest \& Forbes \shortcite{Priest} reproduced the original analysis
by Furth et al.\shortcite{Furth} in a somewhat more concise and
transparent way.) Thanks to this, the nonrelativistic theory and the
relativistic theory should deliver exactly the same results.  At this
point we could simply refer the reader to the studies cited above but
for the sake of completeness we briefly outline the analysis of Priest
\& Forbes \shortcite{Priest}, clarifying few important issues along
the way.  Moreover, in our version we will use the steady state
solution given by eqs.(\ref{b-steady},\ref{v-steady}) instead of the
step function
$$ B_x = \left\{ \begin{array}{lll} +B_0 & \mbox{for} & x>l,\\ B_0 x &
   \mbox{for} & |x|<l,\\ -B_0 & \mbox{for} & x<-l.
\end{array} \right. 
$$ used by the cited authors for the sake of simplicity.

Equation (\ref{vort-lin}) is singular at the point where the wave
vector is normal to the field direction in the equilibrium solution,
that is where $\spr{k}{B}=0$. At this location there develops a thin
{\it resistive sub-layer} where the second order derivatives become
large just like in the classical boundary layer problem (e.g. Bush,
1992). With some loss of generality we will assume in what follows
that $k_z=0$, so that $\spr{k}{B}=kB_x$.  This puts the sub-layer
right in the middle of the current sheet and makes the problem
symmetric with respect to the plane $y=0$.
 
Using $l$ as a unit of length, $\tau_d=l^2/\eta$ as a unit of time, and
$B_0$ as a unit of magnetic field strength we can write the
dimensionless version of equations (\ref{ind-lin},\ref{vort-lin}) as
                                                                                
\begin{equation}
  \dw \db_y = -\dv_y f + (\db_y^{''}-\dk^2\db_y ),
\label{ind-lin1}
\end{equation}
                                                                                
\begin{equation}
\delta (\dv_y^{''}-\dk^2\dv_y)= f \left[-\db_y \frac{f^{''}}{f} +
  (\db_y^{''}-\dk^2\db_y) \right],
\label{vort-lin1}
\end{equation}
where
$$ \db=\frac{\tb}{B_0}, \quad \dk=kl, \quad \dw=\frac{wl^2}{\eta},
\quad \dv_y=-i\tv_y\frac{kl^2}{\eta},
$$
$$ f(\dy)=\frac{B_x}{B_0}=\tanh{\dy}, \quad \dy=y/l,
$$ and
$$ \delta =\frac{\dw}{L_u^2 \dk^2} \ll 1
$$ is a small parameter.  Expanding $\db_y$ and $\dv_y$ in powers of
$\delta$ we find the the leading terms satisfy
                                                                             
\begin{equation}
     \db_y^{''}-\dk^2\db_y +2(1-\tanh^2\dy)\db_y=0,
\label{vort-lin2}
\end{equation}
                                                                             
\begin{equation}
  \dv_y = \frac{\db_y}{\tanh\dy}(2\tanh^2\dy-2 -\dw),
\label{ind-lin2}
\end{equation}
The solution to equation (\ref{vort-lin2}) that satisfies the boundary
condition $\db_y(+\infty) = 0$ is relatively simple

\begin{equation}
  \db_y = \frac{b_0 (\tanh\dy +\dk)}{\dk} \left(
          \frac{1-\tanh\dy}{1+\tanh\dy} \right)^{\dk/2}.
\label{ext-sol}
\end{equation}
Given $\db_y$ one can find $\dv_y$ from eq.(\ref{ind-lin1}).  Once
$b_y$ and $v_y$ are known one can find $b_x$ and $v_x$ from the
conditions
\begin{equation}
  \vdiv{b}=0, \quad \vdiv{v}=0.
\end{equation}

From eq.(\ref{ext-sol}) we find that $\db(0) = b_0$ and then
eq.(\ref{ind-lin1}) shows that $\dv_y$ diverges at $\dy=0$. This
indicates that around $\dy=0$ there exists a sub-layer where the
second order derivatives of $\dv_y$ in equation (\ref{ind-lin}) become
large and cannot be dropped. Thus, solution (\ref{ext-sol}) applies
only to the outside of this sub-layer and is called {\it the outer
solution}. In fact, it holds only for $\dy>0$. The solution that
applies for both sides of the sub-layer is

\begin{equation}
  \db_y = \frac{b_0 (\tanh|\dy| +\dk)}{\dk} \left(
          \frac{1-\tanh|\dy|}{1+\tanh|\dy|} \right)^{\dk/2}.
\label{ext-sol1}
\end{equation}
This solution is continuous at $y=0$ but its first derivative changes
sign at this point so that

\begin{equation}
\Delta'=\left[ \frac{\db'_y}{\db_y}\right]^{0+}_{0-} =
2\frac{1-\dk^2}{\dk}
\label{jump}
\end{equation}
These derivatives are to be used for matching the outer solution with
{\it the inner solution} that describes the interior of the resistive
sub-layer.

To find the structure of the resistive sub-layer, or the inner
solution, we need to introduce new stretched variable $\xi =
\dy/\delta^p$, where the power of $\delta$ is to be found using {\it the principle
of least degeneracy} \cite{VanDyke}. Once $p$ is known we will find
thickness of the sub-layer as $\epsilon = \delta^p$. Here we may
assume that inside the sublayer $\dy \ll1$ and use the approximation
$f=\dy$.  Combining equations (\ref{ind-lin1}) and (\ref{vort-lin1}) we
find
\begin{equation}
  \db_y=\frac{\delta}{\dw}\frac{\dv_y^{''}-\dk^2 \dv_y}{\dy} +
        \frac{\dy}{\dw} \dv_y.
\end{equation}
Substituting this result into eq.(\ref{ind-lin1}) we end up with a 4th
order ODE for $\dv_y$. When using the stretched variable this equation
reads
\begin{equation}
\begin{array}{ll}
  \xi^2\dv^{(4)}-2\dv^{(3)} +& \delta^{4p-1}\dv^{(2)} + 2\delta^{4p-1}
  \dv^{(1)} \\ & - \delta^{6p-1} \xi^4 (\dk^2+2\dw) v = A(\xi,v)
\end{array}
\end{equation}
where $A(\xi,v)$ includes all the term with positive power of
$\delta$.  One can see that when $\dv_y$ is expanded in powers of
$\delta$ the leading order equation will retain the largest number of
terms if $p=1/4$.  Thus the dimensionless thickness of the sub-layer
is
\begin{equation}
   \epsilon = \left(\frac{\dw}{L_u^2\dk^2}\right)^{1/4}
\end{equation}
 
When using the stretch variable eq.(\ref{ind-lin1}) reads
\begin{equation}
  \db_y^{(2)}-\epsilon^2 (\dk^2+\dw)\db_y -\epsilon^3\xi\dv_y=0.
\label{ind-stretched}
\end{equation}
To the second order in $\epsilon$ we can ignore the last term in this
equation and write
\begin{equation}
  \db_y^{(2)}-\epsilon^2 (\dk^2+\dw)\db_y =0.
\label{ind-stretched1}
\end{equation}
The solution of this equation that is symmetric with respect to the
$\xi \to -\xi$ transformation is
$$ \db_y = b_0 \cosh(\sqrt{\dk^2+\dw}\epsilon\xi) .
$$ this gives us the jump in derivatives of $\db_y$ across the
sub-layer
\begin{equation}
 \Delta' = \left[\frac{\db_y^{'}}{\db_y}\right]^{\dy=\epsilon}_{\dy=-\epsilon} =
 \frac{1}{\epsilon}\left[\frac{\db_y^{(1)}}{\db_y}\right]^{\xi=1}_{\xi=-1}
 = 2(\dk^2+\dw)\epsilon.
\label{jump1}
\end{equation}
Here $\db_y^{'}$ is the derivative with respect to $\dy$.

Matching $\Delta'$s derived from the outer and the inner solutions,
eq.(\ref{jump1}) and eq.(\ref{jump}) respectively, allows us to find
the growth rate for the mode with wavenumber $\dk$:

\begin{equation}
  (\dk^2+\dw) \left( \frac{\dw}{L_u^2\dk^2} \right)^{1/4} =
  \frac{1-\dk^2}{\dk}.
\label{g-rate}
\end{equation}
One can see that only modes with wave numbers
\begin{equation}
\dk<1
\end{equation}
grow in amplitude.  For $\dk \ll 1$ dispersion relation simplifies
resulting in the well know equation
\begin{equation}
  \dw = \left(\frac{L_u}{\dk}\right)^{2/5}.
\end{equation}

According to the last result the growth rate increases indefinitely
with the wavelength. However, the e-folding time of the growing mode
\begin{equation}
 \bar{\tau}_e = \frac{1}{\dw}
\end{equation}
cannot be shorter than the resistive time scale of the sub-layer
\begin{equation}
 \bar{\tau}_\epsilon = \frac{\tau_\epsilon\eta}{l^2} = \epsilon^2.
\end{equation}
This condition reads us
\begin{equation}
 \dk > \dk^* = L_u^{-1/4}.
\end{equation}
For smaller $\dk$ the growth is reduced and thus the maximum growth
rate should be of order
\begin{equation}
 \dw^* = \dw_{\dk^*} = L_u^{1/2}
\end{equation}
that corresponds to the dimensional time-scale
\begin{equation}
 \tau^* = (\tau_c \tau_d)^{1/2}.
\end{equation}

\section{Numerical method}
\label{method}

To verify the predictions of the theory of the tearing mode
instability in relativistic magnetically dominated plasma we have
carried out a series of numerical simulations. The equations that are
solved numerically are the Maxwell equations (\ref{Max1}-\ref{Max4})
with the Ohm law
                                                                                
\begin{equation}
\bj = \varrho \frac{\vpr{E}{B}}{B^2}c + \sigma_\parallel \bE_\parallel
 + \sigma_\perp \bE_\perp.
\label{Ohm1}
\end{equation}
This differs from eq.(\ref{Ohm}) by the presence of the conductivity
current across the magnetic field. However, we set $\sigma_\perp$ to
$0$ when $E^2<B^2$ and thus eq.(\ref{Ohm1}) reduces to eq.(\ref{Ohm})
during the linear and the beginning of the non-linear phases of the
tearing instability. However, at some point of the nonlinear phase a
region develops where the electric field becomes stronger than the
magnetic one thus allowing significant cross-field conductivity. We
expect the particle pressure and inertia to become important in such
regions resulting in a different form of the Ohm law. In fact, the MHD
approximation may provide a more suitable mathematical framework
there. However, in this study we are more focused on the linear theory
and for this reason the prescription (\ref{Ohm1}) suffice.  The
numerical scheme used for simulations is a derivative of the general
relativistic scheme for resistive magnetodynamics described in
Komissarov \shortcite{Kom04}. Here, we give its brief description.

From eq.(\ref{Max1}) one finds
\begin{equation}
  \partial_t(\vdiv{B}) = 0.
\label{inforcing}
\end{equation}
This well known result shows that it is sufficient to enforce the
divergence free condition (\ref{Max2}) only for the initial solution
and it will be satisfied automatically at any other $t$.
Unfortunately, straightforward application of many numerical schemes
perfectly suitable for other hyperbolic systems of conservation laws
fails to deliver a good result for electrodynamics and MHD simply
because their discrete equations are not consistent with any discrete
analogue of (\ref{inforcing}).  In particular, this applies to the
method of Godunov which has many beneficial properties and is
currently considered as generally superior to many other numerical
schemes for hyperbolic systems. There have been many attempts to find
a cure for this ``div-B problem'' ( see the review in Dedner et
al.,2002.)
                                                                                  
One of the ways to handle this problem involves construction of a
somewhat different system of differential equations, the ``augmented
system'', where the divergence free condition (\ref{Max2}) is no
longer included and $\vdiv{B}$ may be transported and/or dissipated
like other dynamical variables. The idea is not to enforce the
divergence free condition exactly but to promote a natural evolution
of the system toward a divergence free state. Provided the augmented
system is hyperbolic it can be solved numerically using the Godunov
method \cite{Mun,Ded}.
                                                                                  
Following this idea we modify eq.(\ref{Max1}) as follows
\begin{equation}
  \frac{1}{c}\pder{\bB}{t} + \vcurl{E} + \nabla\Psi =0
\label{Max1-mod}
\end{equation}
where $\Psi$ is the scalar field which is called {\it the
pseudopotential}. It's evolution is given by
\begin{equation}
  -\frac{1}{c}\pder{\Psi}{t} + \vdiv{B}+ \kappa c \Psi = 0.
\label{Psi}
\end{equation}
From these one finds that both $\Psi$ and $\vdiv{B}$ satisfy the
telegraph equation
\begin{equation}
  -\frac{1}{c^2}\frac{\partial^2 \Psi}{\partial t^2} + \kappa
  \pder{\Psi}{t} + \nabla^2 \Psi = 0.
\label{telegraph-P}
\end{equation}
\begin{equation}
  -\frac{1}{c^2}\frac{\partial^2 \vdiv{B}}{\partial t^2} + \kappa
  \pder{\vdiv{B}}{t} + \nabla^2( \vdiv{B}) = 0.
\label{telegraph-B}
\end{equation}

The resultant system includes two vector equations
(\ref{Max1-mod},\ref{Max3}), and one scalar conservation law, equation
(\ref{Psi}). All these evolution equations can be written as
conservation laws. In the form of an abstract vector equation they
read
                                                                                  
\begin{equation}
  \pder{\sqrt{\gamma}{\cal Q}}{t} + \pder{\sqrt{\gamma} {\cal
         F}^{j}}{x^j} = \sqrt{\gamma} {\cal S},
\label{CONS}
\end{equation}
where
                                                                                 
\begin{equation}
  {\cal Q} = \left(\Psi,B^i,E^i \right)
\end{equation}
are the conserved variables
\begin{equation}
   {\cal F}^{j} = c \left(B^j, e^{ijk}E_k+\Psi g^{ij}, - e^{ijk}B_k
   \right)
\end{equation}
are the corresponding hyperbolic fluxes,

\begin{equation}
{\cal S} = (\kappa c^2 \Psi,0^i,4\pi j^i).
\end{equation} 
In this study we employ Cartesian coordinates so that 1) the metric
tensor of space $g^{ij}$ equals to Kronecker's delta $\delta^{ij}$, 2)
its determinant $\gamma=1$, and 3) the Levi-Civita alternating tensor
$e^{ijk}$ reduces to the Levi-Civita symbol with $e^{123}=1$.

Numerical integration of (\ref{CONS}) is carried out using a second
order Godunov method based in combination with the time-step splitting
technique by Strang \shortcite{Str} for handling the source term.
Namely, each time-step $t^n \rightarrow t^n+\Delta t$ is split into
three sub-steps.  During the first sub-step the numerical solution is
advanced in time by $\Delta t/2$ via integrating the truncated system
\begin{equation}
  \pder{\cal Q}{t} = {\cal S}.
\label{S1}
\end{equation}
In fact we split the source term into two parts
\begin{equation}
            {\cal S}_a =(\kappa c^2\Psi ,0^i,4\pi j_c^i) \text{and}
            {\cal S}_b =(0,0^i,4\pi j_d^i),
\label{Sa}
\end{equation}
where $\bj_c=\sigma_\parallel \bE_\parallel + \sigma_\perp \bE_\perp$
is the conductivity current and $\bj_d = \varrho c\vpr{E}{B}/{B^2}$ is
the drift current, and split this substep as well. First we account
for ${\cal S}_a$ only. This source term is potentially stiff but its
simple form allows exact analytical integration of eq.(\ref{S1}) thus
reducing the stability constraints on the timestep.  Then we integrate
eq.(\ref{S1}) with source term ${\cal S}_b$ numerically using the
method of Newton.  During the second sub-step the resultant solution
is advanced in time by $\Delta t$ via integrating the truncated system
                                                                                
\begin{equation}
  \pder{\sqrt{\gamma}\cal Q}{t} + \pder{\sqrt{\gamma}{{\cal
  F}^j}}{x^j} =0
\label{S2}
\end{equation}
using the second order Godunov scheme. The third sub-step is a repeat
of the first one but now it is the solution found by the end of the
second sub-step that is used as the initial solution for
eq.(\ref{S1}).

In our Godunov scheme the numerical fluxes through the cell interfaces
are found using the exact solution to the interface Riemann
problems. The construction of this exact solver is simplified by the
fact that system (\ref{S2}) is linear. Its 1D-version can be written
as
\begin{equation}
   \pder{\cal Q}{t} + {\cal A} \pder{\cal Q}{x} = 0
\label{1Dsystem}
\end{equation}
with the Jacobean matrix
\[
{\cal A} = c \left[
\begin{array}{ccccccc}
 0 & 1 & 0& 0 & 0 & 0 & 0 \\ 1 & 0 & 0& 0 & 0 & 0 & 0 \\ 0 & 0 & 0& 0
 & 0 & 0 & -1 \\ 0 & 0 & 0& 0 & 0 & 1 & 0 \\ 0 & 0 & 0& 0 & 0 & 0 & 0
 \\ 0 & 0 & 0& 1 & 0 & 0 & 0 \\ 0 & 0 & -1& 0 & 0 & 0 & 0 \\
\end{array}
\right] .
\]
The eigenvalues of this matrix
$$ \mu_{(1)} = \mu_{(2)} = \mu_{(3)} = c;
$$
$$ \mu_{(4)}=0;
$$
$$ \mu_{(5)}=\mu_{(6)}=\mu_{(7)} = -c
$$ provide the wavespeeds of hyperbolic waves. Other properties of
these waves are given by the eigenvectors of ${\cal A}$.  The right
eigenvectors are
                                                                                  
\begin{equation}
\begin{array}{ccc}
\br_{(1)} & = & (1,1,0,0,0,0,0) \\ \br_{(2)} & = & (0,0,-1,0,0,0,1) \\
\br_{(3)} & = & (0,0,0,1,0,1,0) \\ \br_{(4)} & = & (0,0,0,0,1,0,0) \\
\br_{(5)} & = & (-1,1,0,0,0,0,0) \\ \br_{(6)} & = & (0,0,1,0,0,0,1) \\
\br_{(7)} & = & (0,0,0,-1,0,1,0) \\
\end{array}
\end{equation}
Since ${\cal A}$ is symmetric its left eigenvectors, $l_{(i)}$,
coincide with the corresponding right eigenvectors:
\begin{equation}
 l_{(i)} = r_{(i)},
\end{equation}
                                                                                  
It is easy to see that solutions 2,3,6, and 7 are the usual
electromagnetic waves. Solution 4 simply reflects the fact that all
waves of vacuum electrodynamics are transverse and any discontinuity
in the normal component of electric field is due to a surface electric
charge distribution.  Solutions 1 and 5 describe new waves that do not
exist in electrodynamics; they transport $\Psi$ and $\vdiv{B}$.

The solution to the Riemann problem with the left and the right
states, ${\cal Q}_{(l)}$ and ${\cal Q}_{(r)}$, and the interface speed
$\beta^x$ is
\begin{equation}
  {\cal Q} = {\cal Q}_{(l)} + \sum\limits_{i=1,3} \alpha_{(i)}
  \br_{(i)}
\end{equation}
where the wave strengths
$$ \alpha_{(i)} = \frac{({\cal Q}_{(r)}-{\cal Q}_{(l)} )\cdot
  \br_{(i)}} {\br_{(i)}\cdot \br_{(i)}}.
$$ Notice, that the 4th wave is ignored. The x-component of electric
field is set to be the mean value of the left and the right states
                                                                                  
\begin{equation}
 E^x = 0.5 (E^x_{(l)} + E^x_{(r)}) .
\end{equation}

The entire scheme is second order accurate in space and time provided
$\Delta t \leq \eta/c$. The only stability constraint on the time step
comes from the Currant condition $\Delta t < \Delta x/c$ for
eq.(\ref{S2}). However, the accuracy considerations may require a
smaller time step. For example, our 1D test simulations of the
equilibrium current sheet (\ref{b-steady},\ref{v-steady}) have shown
that if $\Delta t = \eta/c$ then the relative error in $E_z$ is around
$10\%$.

\begin{figure*}
\includegraphics[width=75mm]{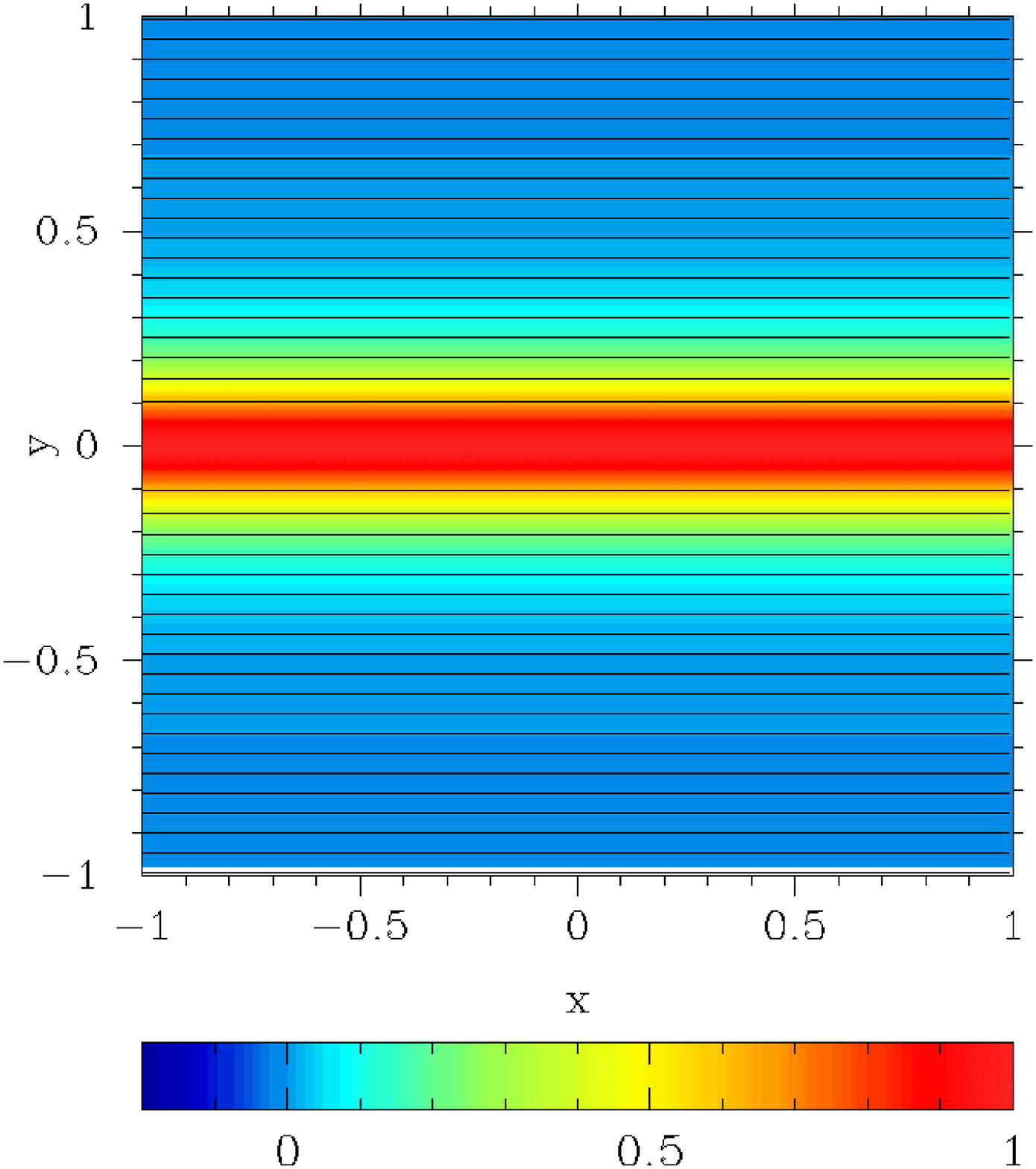}
\includegraphics[width=75mm]{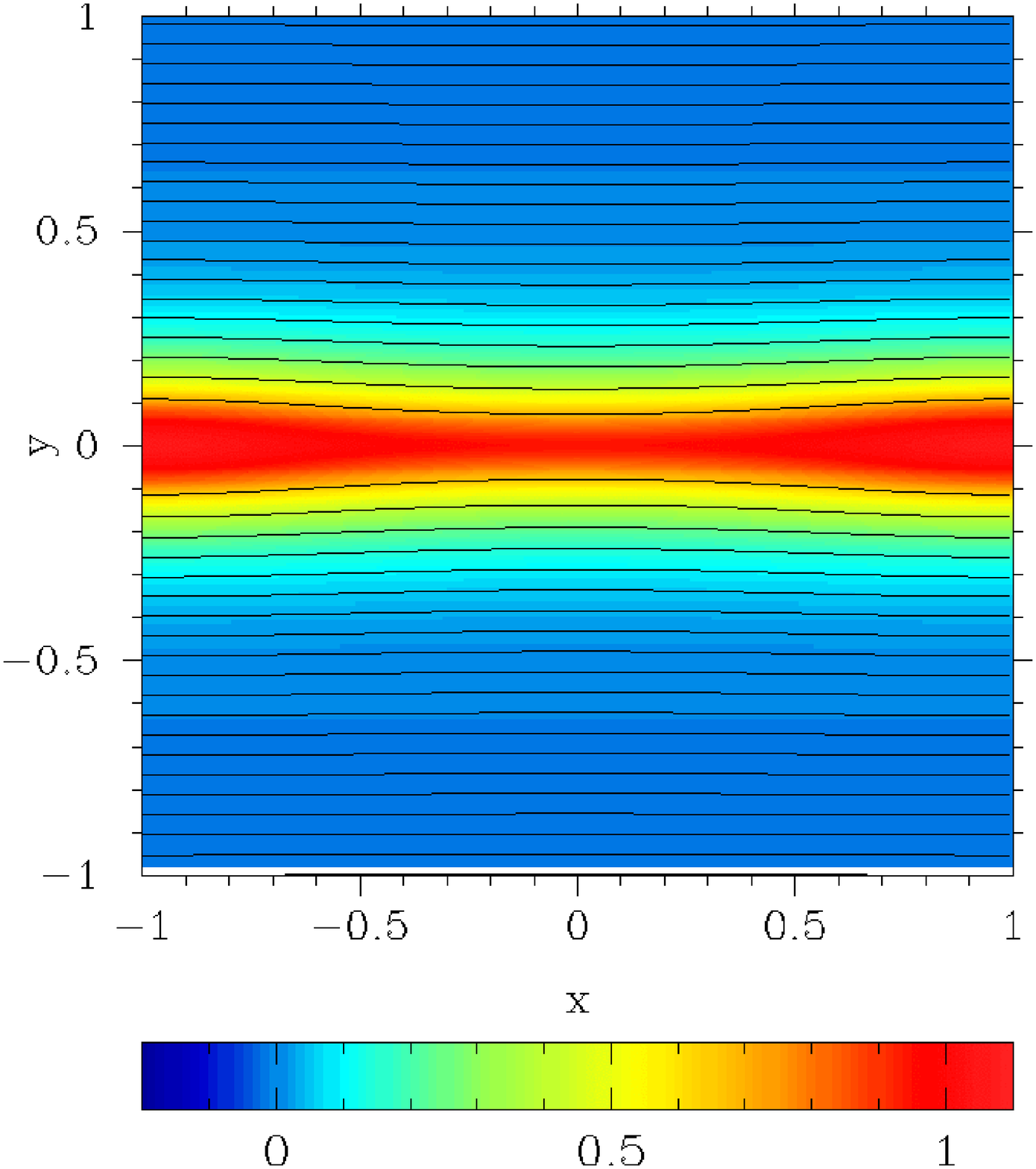}
\includegraphics[width=75mm]{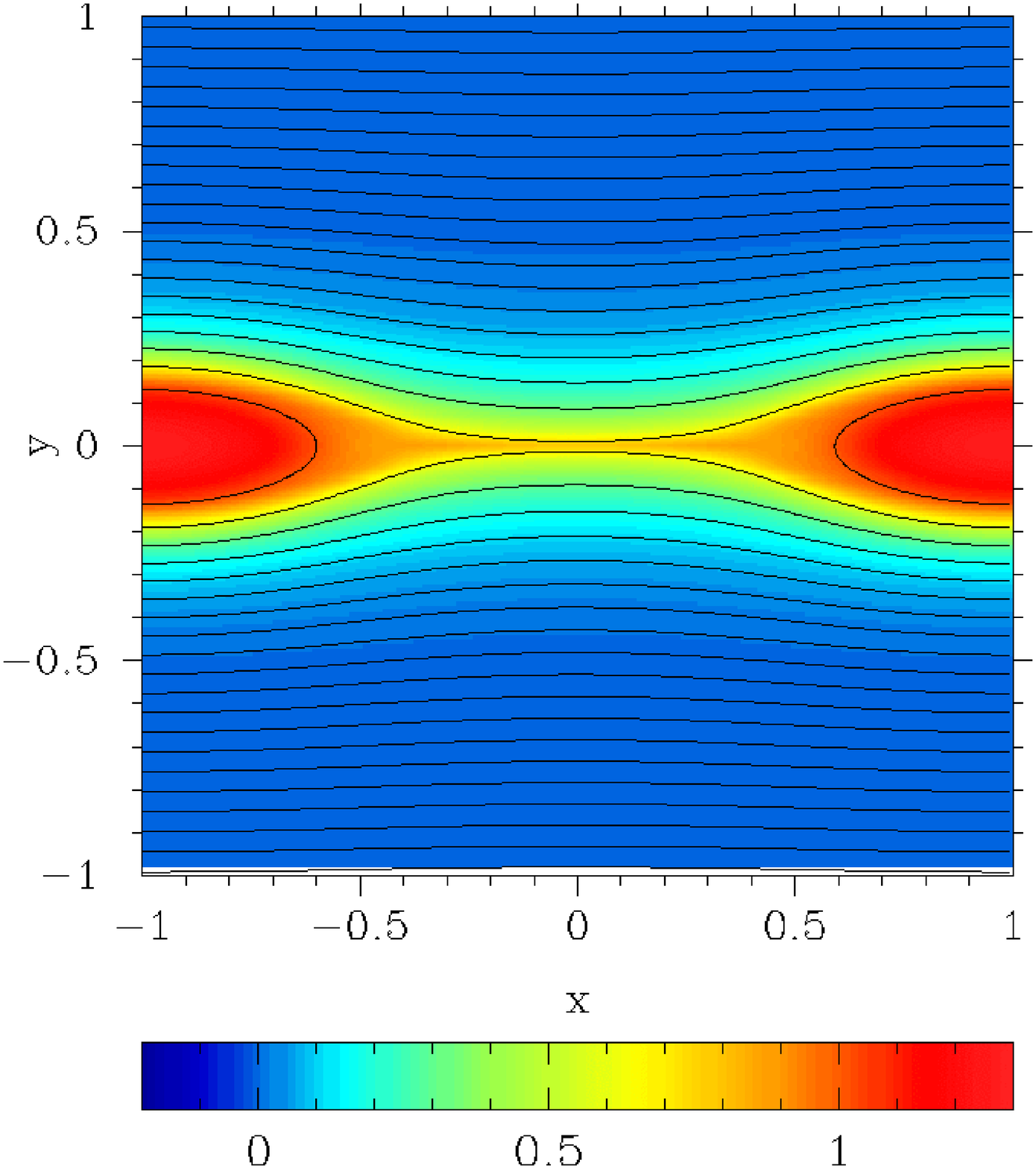}
\includegraphics[width=75mm]{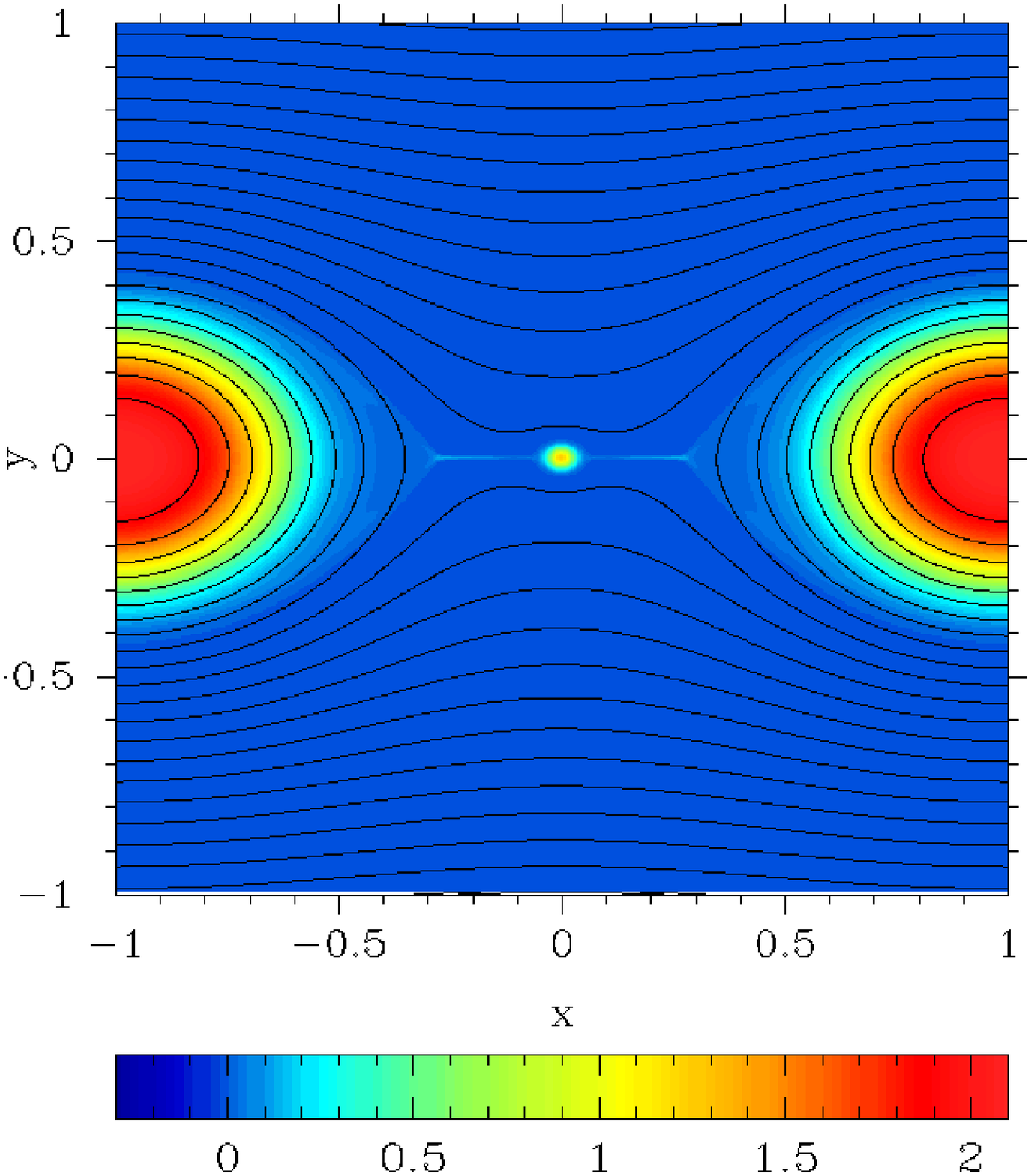}
\caption{The evolution of the current sheet with $\dk=0.314$ ($\dk
L_u^{1/4}=1.08$). $B_z$ (color image) and the magnetic field lines at
$t=0$ (top left), $t=5$ (top right), $t=10$ (bottom left), and $t=16$
(bottom right).  }
\label{evol}
\end{figure*}

\section{Computer Simulations}
\label{simulations}

These are 2D simulations with slab symmetry in the z-direction (that
is $\partial f /\partial z = 0$ for any function $f$). The utilized
units are such that $l=0.1$, $c=1$, and $B_0=1$. Following the setup
of Sec.\ref{tearing} the initial equilibrium current sheet is normal
to the y-axis and its symmetry plane is $y=0$. The computational
domain is $[-x_0,+x_0]\times[-y_0,+y_0]$ with $x_0=\lambda/2$, the
half of the perturbation wavelength, and $y_0=2.0$.  On the boundaries
$x=\pm x_0$ we impose the periodic boundary conditions and on the
boundaries $y=\pm y_0$ we impose the zero-gradient conditions.  In
order to get accurate solutions we have to have the resistive
sub-layer well resolved. On the other hand, far away from the
sub-layer the solution has no fine structure and does not require high
resolution.  This suggests to use a grid with variable resolution in
the y-direction.  We adopted the exponential rule
$$ \Delta y_{i+1} = \left\{
\begin{array}{rl} 
\alpha \Delta y_i & \text{for} y_i<0,\\ \alpha^{-1} \Delta y_i &
\text{for} y_i>0,
\end{array} 
\right.
$$ with the resolution in the symmetry plane being 10 times higher
than that at the domain boundary, $\Delta y(0)=0.1\Delta y(y_0)$. The
resolution in the x-direction was uniform with $\Delta x = \sqrt{0.1}
\Delta y(y_0)$.

\begin{figure}
\includegraphics[width=85mm]{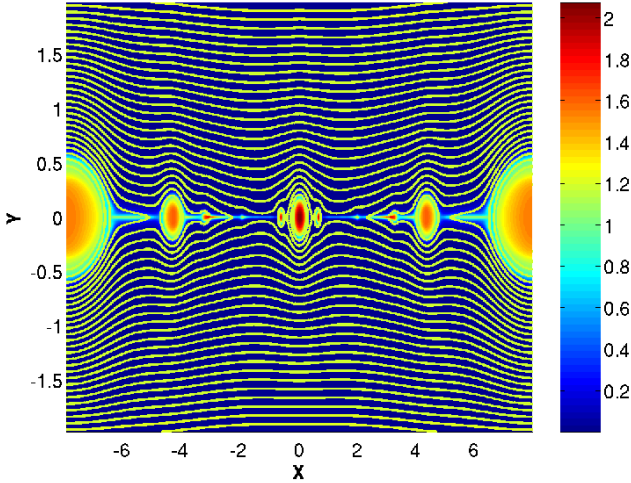}
\caption{Multiple magnetic islands at the nonlinear stage of the model
with $\dk=0.0393$ ($\dk L_u^{1/4}=0.135$, $\eta = 10^{-3}$).}
\label{llong}
\end{figure}

We have tried two different models for the initial solution.  In the
first model the perturbation of the equilibrium solution
(\ref{b-steady},\ref{e-steady}) has the form
$$ \bb=(0,b_0\sin(\pi x/x_0),0), \quad \be =(0,0,0)
$$ with $b_0=10^{-3}$.  Since this is not the normal mode of the
tearing instability the corresponding numerical solutions exhibited
initial settling period of order $\delta t \simeq 1$ before reaching
the stage of exponential growth.

In the second model we tried to set up the normal mode. In fact, we
used the the outer solution (\ref{ext-sol},\ref{ind-lin2}) with
$b_0=10^{-3}$ in order to introduce the perturbation. Since, the
electric field given by this solution diverges at $y_0$ we resorted to
linear interpolation in order to continue it within the resistive
sub-layer. Perhaps this explains why the solution still exhibited
approximately the same settling time as in first model.  Moreover, we
have not found any significant differences between the solutions
corresponding to the both models except from their behavior during the
settling.

We have considered only two values for the resistivity, $\eta=10^{-3}$
$\eta=10^{-4}$, corresponding to the Lundquist number $L_u=1.4\times10^2$ and
$L_u=1.4\times10^3$ respectively.  Figure \ref{evol} shows the evolution of the
current sheet for $\eta=10^{-3}$ and $\dk=0.314$. As the perturbation
grows the current sheet gradually thins out in the middle of the
domain and thickens at the x-boundaries. Eventually this results in
the development of two large magnetic islands. By the end of the
simulations a third much smaller island forms right in the middle. In
fact, such secondary islands are typical and more pronounced for
longer wavelengths.  Figure \ref{llong} shows the final solution for
the model with $\dk=0.039$.  In this case there are free relatively
large secondary islands and few smaller secondary islands in between
them. These secondary islands originate in the following order. The
larger secondaries develop when the two primaries suck in a large
fraction of the current lines so that the residual current sheet
becomes rather thin. These first secondary islands also suck in the
current lines and this leads to further reduction of the thickness of
the residual. Then new smaller islands appear between secondary island
of the first generation and so on.  This behavior seems to be caused
by the reduction of the growth time for the tearing mode in thin
current sheets.

As the thickness of the residual current sheet reduces, $\nabla\times
B$ goes up and so do the electric field and electric current density
in the sheet. At some point the electric field, which is predominantly
perpendicular to the magnetic field, reaches the same strength as the
magnetic one and the drift speed speed reaches the speed of
light. Eventually the current sheet collapses to the size of a
computational cell and its electric field becomes even stronger than
the magnetic field. This shows that the structure of the current sheet
can no longer be described within our model and other factors like the
thermodynamic pressure of plasma heated to high temperatures in the
current sheet has to be taken into account.  Prior to this point the
quasi-equilibrium of the current sheet is basically supported by the
magnetic pressure alone, the pressure of predominately x-directed
magnetic field outside of the current sheet being balanced by the
pressure of predominately z-directed magnetic field inside of it.

\begin{figure}
\includegraphics[width=85mm]{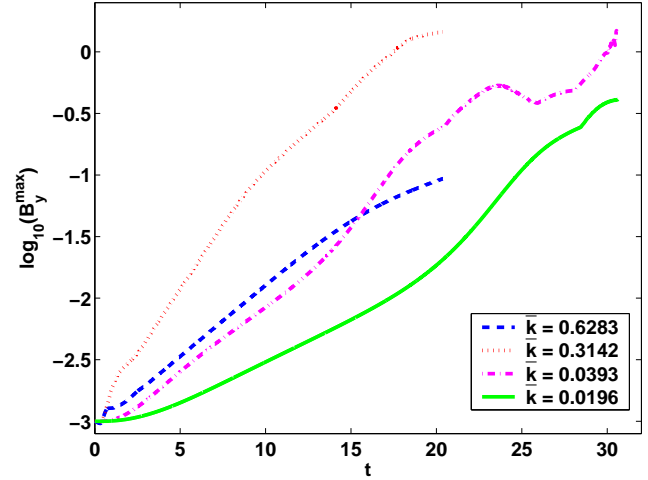}
\caption{ Growth of the perturbation for various wavenumbers ($\eta
= 10^{-3}$). $B_y^{\mbox{\tiny max}}$ is the maximum value
of $B_y$ on the grid. }
\label{grow}
\end{figure}

\begin{figure}
\includegraphics[width=85mm]{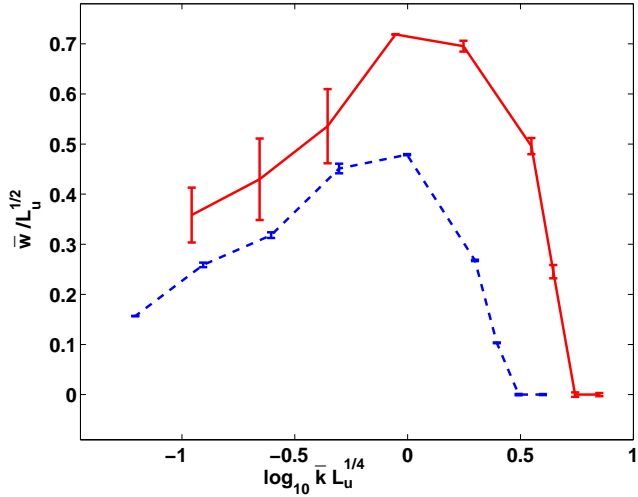}
\caption{ Growth rate as a function of the wavelength.  {\it Dashed
line:} $\eta = 10^{-3}$ {\it solid line:} $\eta =
10^{-4}$.}
\label{disp}
\end{figure}

Figure \ref{grow} shows how the maximum value of $B_y$ grows with time
for different wavelengths of the perturbation in the case of
$\eta=10^{-3}$.  One can clearly see that after the short initial
settling period the growth of $B_y$ becomes exponential. At this phase
the primary islands dominate in the solution.  This phase terminates
when the nonlinear effects become significant or when the smaller
secondary islands overtake the primary ones (see the curves for
$\dk=0.0393$, and 0.0196).  There are two reasons for the faster
growth of the secondary islands.  Firstly, at the point of their
appearance the thickness of the current sheet has been significantly
reduced. Secondly, both the primary and the secondary islands
correspond to wavenumbers $\dk < \dk^*$.  In this regime (see figure
\ref{disp}) the perturbations of shorter wavelengths grow faster.

Figure \ref{disp} shows the dispersion relations $\dw=\dw(\dk)$ that
accounts for the growth rates of the primary islands.  One can see
that the dispersion curves have maximum at around $\dk^* = L_u^{-1/4}$
that agrees very well with the analytical results presented in
Sec.\ref{tearing}. The maximum growth rate is also close to the
predicted $\dw_{\dk^*}=L_u^{1/2}$. Moreover, the results clearly
indicated that modes with $\dk>\dk_c=1$ are indeed stable (see
eq.\ref{g-rate}).  The reason why the results for $\eta=10^{-4}$
give the maximum growth rate somewhat closer to the
theoretical value than those for $\eta=10^{-3}$ may have
to do with the fact that the separation between the cut-off
wavenumber, $\dk_c$, and $\dk^*$ becomes too short for relatively 
low Lundquist numbers.

 \section{Application to magnetar flares}
 \label{applic}

One of the main unknowns in our calculations is the value of the
resistivity $\eta$, which should be calculated from the particle
kinetics, but instead was introduced as a macroscopic property of
plasma - a common approach in continuous mechanics.  Due to very short
radiative decay times in magnetar magnetospheres the particles are
bound to move only along the field lines, this limits considerably a
number of possible resonant wave-particle interactions that can lead
to development of plasma turbulence.  The main remaining options are
the Langmuir turbulence, which in relativistic plasma develops on a
typical scale of electron skip depth, $\delta_e \sim c/\om_{p,e}$, and
ion sound turbulence, which in relativistic plasma develops on an ion
skip depth $\delta_i \sim c/\om_{p,i}$ ($\om_{p,e}$ and $\om_{p,i}$
are the electron and ion plasma plasma frequencies). They are
different by a square root of the ratio of electron to ion masses
$\delta_e/\delta_i \sim (m_e /m_i)^{1/2}$.  A fully developed
turbulence with a typical velocity $c$ and typical scale $\delta$
would produce a resistivity $\eta \sim c \delta$. Thus we can write
the growth time of tearing mode as \beq \tau \sim {l \over c} \left( {
l\over\delta} \right)^{1/2} \ee Notice, that the result does not
depend very much on which of the skin depths we use.  The difference
is in the factor $ (m_e /m_i)^{1/4} \simeq$ few.  In current-carrying
magnetospheres of magnetars the plasma frequency can be estimated as $
\om_p = \lambda \sqrt{ \om_B r/c} $ ($\lambda $ is a parametrization
constant and $r$ is the radial distance from the neutron star center)
giving a growth time of the tearing mode \beq \tau = 2 (B/B_Q)^{1/4}
(l/R_{NS})^{7/4} \sqrt{\lambda}\,\, \mbox{msec},
\label{tau}
\ee where $R_{NS}$ is the neutron star radius and $B_Q$ is the quantum
magnetic field.  We also note that the expected Lundquist number is
$L_u \sim l/\delta \sim 5 \times 10^3 (B/B_Q)^{1/2} (l/R_{NS})^{3/2}
\lambda ^{-1/2}$, which is close to the values used in our numerical
simulations.

Although the connection between the growth time of the tearing mode
and the observed flare rise time may not be that straightforward, the
fact that the estimate (\ref{tau}) is close to a typical rise time of
magnetar flares \cite{gogus} is encouraging.  Keeping in mind the
uncertainties in the estimates of $\eta$ and $l$ we conclude that our
model is consistent with observations.

 \section{Conclusions} 

The original idea of this study was to verify the surprising
analytical results of \cite{Lyu03} on the tearing instability in
magnetically dominated relativistic plasma by means of direct
numerical simulations. We have carried out such simulations and they
fully confirmed Lyutikov's conclusion on the growth rate in
magnetically dominated regime -- it indeed coincides with that found
earlier for the tearing instability in non-relativistic incompressible
MHD, the shortest growth time being equal to the geometric mean of the 
\Alfven and resistive
timescales \cite{Furth,Priest}.  Moreover, like in the
non-relativistic case we observed formation of magnetic islands in
the non-linear phase of the instability. This is accompanied by the
appearance of noticeable O-type and X-type neutral points (or
lines). The development of an X-point may be considered as a first
step in setting up of reconnection layer where dissipated magnetic
energy and magnetic flux are constantly replenished due to plasma
inflow \citep{syr81}.
                                                                                          
Our results support the possibility of magnetar flares being 
magnetically driven reconnection-like events that occur in magnetar
magnetospheres \citep{Lyu03,Lyu06} in a similar fashion to flares 
and coronal mass ejections of Sun's magnetosphere.  
More complicated \Bf geometries, e.g. those with nonvanishing magnetic 
tension, should also be subject to tearing instability on time scales intermediate 
between the resistive and \Alfven time scales \citep{Priest}. The growth rate
typically scales as $\tau \sim \tau_c^{\alpha} \tau_d^{1-\alpha}$, where
$0<\alpha<1$ is some coefficient ($\alpha=1/2$ in our case).  In the 
case of magnetar flares the observed flare rise-times range from fractions of
a millisecond to $\sim $ ten milliseconds, which is indeed intermediate 
between the \Alfven crossing time, $\sim 0.03 $ msec, and the resistive time, 
$\sim$ seconds, of the magnetosphere \cite{Lyu03}.

The short timescale of tearing mode is achieved entirely through
the formation of a very narrow resistive sub-layer with very short
diffusion time scale. However, it is also expected that the
resistivity itself may be enhanced inside such sub-layers due to the
development of plasma turbulence (anomalous resistivity).  Elucidating
the properties of anomalous resistivity in such plasma is an important
next step. Moreover, the collapse of current sheets and the eventual
development in our simulations of the drift speeds exceeding the speed
of light clearly indicates a breakdown of the MD approximation. In
order to move forward one needs to put back the dynamical
effects of matter which are bound to become important in this singular
domain. The simplest way of doing this is to return to the model of
relativistic MHD like in Lyutikov \& Uzdensky \shortcite{lu03} and
Lyubarsky \shortcite{Lyub05}.

The coincidence of the growth rates for the tearing instability in
incompressible MHD and MD clearly suggests some hidden similarity
between the evolution equations of these two systems. We have carried
out further analysis of resistive MD and discovered that in the
quasi-equilibrium limit, that is characterized by low drift speeds, its
equations do indeed become similar to those of non-relativistic
MHD. In particular, the mass-energy density of the magnetic field takes on the role
the mass density of classic MHD whereas the drift velocity plays the
role of the fluid velocity. {\it So the dynamics of the system can be
described as a flow of magnetic mass-energy under the action of
magnetic pressure and tension.}  Moreover, the fact that the magnetic
tension of the equilibrium current sheet subjected to tearing
instability is zero allows further reduction of the MD equations which
leads to almost the same system as incompressible MHD.  When, this
system is utilized in the analysis of the tearing instability it
generates exactly the same equations for the perturbation as in
incompressible MHD. Thus, our analysis provides perfect mathematical
explanation for the surprising results in Lyutikov \shortcite{Lyu03}
and improves our understanding of the dynamics of magnetically-dominated
plasma.

Finally, we seem to have come up with a more suitable name for the
system of equations describing the dynamics of relativistic
magnetically dominated plasma. Instead of the currently used {\it
force-free degenerate electrodynamics} (FFDE) or simply {\it
force-free electrodynamics} (FFE) we propose to use {\it
magnetodynamics} (MD).  Let us summarize our arguments in favour of
this new name. The current name originates from the early studies of
stationary magnetospheres of neutron stars and black holes.  Besides
the steady-state Maxwell's equations, the key equations used in those
studies were the degeneracy condition, $\spr{E}{B}=0$, and the
condition of vanishing Lorentz force. The full system of the
corresponding time-dependent equations was not known at the time and
so were not known its connection with the relativistic MHD and the
properties of it's waves.  All these have been discovered only very
recently \cite{Uchida,Gru99,Kom02}.  In Komissarov \shortcite{Kom02}
the dynamical equations of relativistic magnetically dominated plasma,
equations (\ref{MD1}-\ref{MD4}), have been derived from the system of
ideal relativistic MHD in the limit of vanishing rest mass density and
pressure of matter.  This way of derivation immediately suggests
simply to remove the {\it hydro}-component from the word {\it
magneto-hydro-dynamics}, which obviously results in the name we
propose here. Like in ideal MHD, in ideal MD the electric field also
vanishes in the the fluid frame, or to be more precise in the frame
moving with the drift velocity.  An inertial observer moving with this
velocity would experience a pure magnetic field.  Thus, it is
desirable to retain the emphasis on the magnetic component of the
electromagnetic field, that is present in the name of MHD.  The key
wave of MHD, the Alfv\'en wave, is also present in MD. It propagates
along the magnetic field lines with the speed of light \cite{Kom02}.
Like in MHD the electric field vector can be eliminated from the set
of dependent variables of equations (\ref{MD1}-\ref{MD4}) (see
Sec.\ref{basics}).  Finally, as we have shown in this paper one can
qualitatively describe the evolution of magnetically dominated plasma
as a flow of magnetic energy under the action of Maxwell's
stress. Apparently, we are not the only ones who are not particularly
happy with the name force-free electrodynamics. Recently, A.Spitkovsky
proposed to use the name {\it force-free MHD} instead \cite{Spi06}.
This name is somewhat better but it is still not completely
satisfactory. First of all, matter and the electromagnetic field
appear in this name on equal terms, contrary to the relevant physical
conditions. Secondly, the name MHD implies a somewhat different set of
evolution equations. Thirdly, as we have seen, the resistive case can
no longer be described as force-free. Finally, the name {\it
magnetodynamics} is made of just one word that makes it more
esthetically pleasing.

\section*{Acknowledgments}
 This research was funded by PPARC under the rolling grant
``Theoretical Astrophysics in Leeds''


\end{document}